\documentclass[aps,twocolumn, preprintnumbers, showpacs, prl]{revtex4}
\usepackage{graphicx}
\usepackage{times}
\usepackage{epsfig}
\usepackage{amsmath}
\usepackage{amsfonts}
\usepackage{amssymb}
\usepackage{url}
\usepackage{hyperref}
\usepackage{subfigure}
\newcommand{\beqa}{\begin{eqnarray}} 
\newcommand{\eeqa}{\end{eqnarray}} 
\newcommand{\beq}{\begin{equation}} 
\newcommand{\eeq}{\end{equation}}

\newcommand{\bmt}{\begin{pmatrix}}
\newcommand{\emt}{\end{pmatrix}}
\newcommand{\be}{\begin{equation}}
\newcommand{\ee}{\end{equation}}
\newcommand{\bea}{\begin{eqnarray}}
\newcommand{\eea}{\end{eqnarray}}

\begin{document}
\title{$7$ keV sterile neutrino Dark Matter in extended seesaw framework}
\author{Sudhanwa Patra}
\email{sudha.astro@gmail.com}
\author{Prativa Pritimita}
\email{pratibha.pritimita@gmail.com}
\affiliation{Centre of Excellence in Theoretical and Mathematical Sciences, Siksha \textquoteleft O\textquoteright 
             Anusandhan University, Bhubaneswar, India}
\begin{abstract}
 The appearance of a $3.5$ keV photon line in the X-ray spectra of Andromeda galaxy and other galaxy clusters 
including the Perseus galaxy can be interpreted as the signal of a $7$ keV sterile neutrino dark matter 
candidate as one of the plausible explanations. We present here a novel framework 
of extended seesaw mechanism where the lightest sterile neutrino acts as a Dark Matter 
candidate and its decay into a photon and an ordinary neutrino accounts for  
the excess 3.5 keV X-ray line. For implementing the idea, we add two different types of neutral fermion singlets 
($S_L$, $N_R$) to the SM light neutrino $\nu_L$ so that the light neutrino mass 
is governed by generic linear seesaw formula. We discuss the complementarity constraints 
on the model parameters including sterile-active neutrino mixing angle within the extended 
seesaw framework. Favorably, this framework can be embedded in a $SO(10)$ grand unified theory.
\end{abstract}
\maketitle
\section{Introduction}
 Dark Matter (DM), which occupies $25\%$ of the total energy density of our present 
universe, still remains a mystery from particle physics point of view. Though its exploration 
seems quite difficult with a limited knowledge about its properties, a pretty good number of papers 
including ref.\cite{dm} have brought the issue little closer to our reach. The observation of an excess 3.5 keV 
photon line in the X-ray spectra of Andromeda galaxy and many other galaxy clusters by XMM-Newton X-ray 
Space observatory \cite{XMM-Newton-a,XMM-Newton-b} has made the topic afresh since the possibility of 
its appearance due to the decay of DM can not be exactly denied. This weak line has created much noise 
in the astro-particle world recently with the observed flux and best fit energy peak being at, 
\begin{eqnarray}
&& \Phi_{\gamma} = 4 \pm 0.8 \times 10^{-6}\, \mbox{photons \,cm}^{-2} \mbox{sec}^{-1} \nonumber \\
&& E_{\gamma} = 3.57 \pm 0.02 \, \mbox{keV}
\end{eqnarray}

 The aftermath of this finding brings Dark Matter into the plot since no other atomic transition in 
thermal plasma can account for this intensity. The potential candidates responsible for giving such a 
result can be a sterile neutrino dark matter 
\cite{DM-decay-a,DM-decay-b,DM-decay-c,DM-decay-d,DM-decay-e,DM-decay-f,DM-decay-g,DM-decay-h}, axions 
or axion like particles \cite{DM-axion-a,DM-axion-b,DM-axion-c,DM-axion-d,DM-axion-e}, 
axinos \cite{DM-axinos-a,DM-axinos-b,DM-axinos-c}, 
gravitino as decaying dark matter \cite{DM-susy-a,DM-susy-b,DM-susy-c}, moduli \cite{DM-moduli-a,DM-moduli-b,DM-moduli-c}, 
Magnetic dipolar dark matter \cite{DM-dipole-a,DM-dipole-b,DM-dipole-c,DM-dipole-d,DM-dipole-e,DM-dipole-f,DM-dipole-g}, 
and others \cite{DM-others-a,DM-others-b,DM-others-c}.

We humbly look for a possible explanation of the origin of this monochromatic 3.5 keV photon line 
with a 7 keV sterile neutrino dark matter as the prime tool within an extended seesaw framework. The 
decay of this sterile neutrino dark matter into a photon and an ordinary neutrino can explain the 
appearance of the X-ray signal with energy $3.51$ keV. The paper is sketched as follows. 
In Sec.II, we discuss briefly the extended linear seesaw framework and provide correct picture 
of masses and mixing with the ordinary neutrinos in the one generation case. In Sec.III, 
we extend the study in a three generation picture using the consistent 
model parameters resulting lightest sterile neutrino mass around $7$ keV and active-sterile 
mixing angle around $10^{-5}$. We present the status of keV sterile neutrino dark matter 
including the stability issue and relic abundance in Sec.IV with the existing experimental 
constraints. Towards the end, we summarize our results and conclude the work.
\section{Framework accommodating $7$ keV Dark Matter}
We have added two extra neutral fermions $N_i, S_i\, \mbox{i=1,2,3}$ 
that are singlets under SM gauge group $SU(2)_L \times U(1)_Y$ to the usual SM leptons
$$\ell_L=\begin{pmatrix}
         \nu_\alpha \\ e_\alpha 
         \end{pmatrix}_L\, , \quad e_{\alpha R}\, ,
$$
where $\alpha=e,\mu,\tau$ represents a flavour/generation index. The extra fermions 
being neutral and singlets under SM gauge group, can have Majorana mass terms 
(i.e, $M_R$ for $N$ and $\mu$ for $S$) in the interaction Lagrangian. 
A notable feature of the model is that, the mass term for RH heavy neutrino ($M_R$) is purposely larger 
than other mass terms present in the model.

The extended seesaw mechanism featuring similar mass matrices for Dirac neutrino and up-quark can give 
interesting results since type I seesaw contribution is exactly canceled out here and can be accessible to LHC. 
Thus we have chosen here an extended Linear seesaw framework woven with the addition of a large Lepton number 
violating mass term $M_R$ to the minimal linear seesaw scheme \cite{ext-seesaw-a,ext-seesaw-b,ext-seesaw-c,
ext-seesaw-d,ext-seesaw-e,ext-seesaw-f,ext-seesaw-g} where the lightest sterile neutrino $S_{\rm lightest}$ 
acts as a Dark Matter candidate. We present below the interaction Lagrangian, masses and mixing 
of the framed model and examine whether it results $7$ keV sterile neutrino dark matter with adequate mixing 
with the active neutrinos to be able to account for the unidentified $3.5$ keV X-ray line.

The relevant interaction Lagrangian for extended linear seesaw framework having both the 
lepton number violating Majorana mass terms for $N$ and $S$ with $\mu=0$ is as follows,
\begin{eqnarray}
-\mathcal{L}_{\rm mass} &= & 
                            M_D \overline{\nu_L} N_R 
                          + M_L \overline{(\nu_L)^C} S_L \nonumber \\
                       && + M \overline{S_L} N_R + \frac{1}{2} M_R \overline{N^c_R} N_R \text{+ h.c.}
\end{eqnarray}
where $M_D$ is the Dirac neutrino mass matrix connecting $\nu_L-N_R$, $M$ is the heavy $N-S$ mixing matrix, 
$M_L$ is the lepton number violating $\nu-S$ mixing matrix and $M_R$ is the large Majorana mass matrix for 
heavy RH neutrinos $N_R$. After the electroweak symmetry breaking, one can write down the full ($9 \times 9$) 
neutrino mass matrix in the ($\nu_L$, $N_R$, $S_L$) basis as,
\begin{equation}
\mathcal{M}= \left( \begin{array}{ccc}
                0        & M_D  & M_L   \\
                M^T_D    & M_R  & M^T \\
                M^T_L    & M  & 0
                      \end{array} \right) \, ,
\label{eqn:numatrix-linear}       
\end{equation}
The allowed hierarchy between different neutral fermion 
mass matrices is $M_R > M, M_D \gg M_L$. Since the lepton number violating term for 
heavy RH neutrinos is heavier than the other mass scales, it is eventually integrated out 
from the Lagrangian resulting in
\begin{eqnarray}
&&\hspace*{-1.0cm}- \mathcal{L}_{\rm eff} = \left(M_D \frac{1}{M_R} M^T_D 
          - M_D M^{-1} M^T_L -M_L M^{-1} M_D^T\right)_{\alpha \beta}\, \nu^T_\alpha \nu_\beta \nonumber \\
&&\hspace*{1cm}+
\left(M_D \frac{1}{M_R} M^T \right)_{\alpha j}\, \left(\overline{\nu_\alpha} S_j + \overline{S_j} \nu_\alpha \right) \nonumber \\
&&\hspace*{1cm}+\left(M \frac{1}{M_R} M^T\right)_{ij}\, S^T_i S_j \, ,
\end{eqnarray}
which, in the $\left(\nu, s\right)$ basis, gives the $6 \times 6$ mass matrix
\begin{eqnarray}
&&\mathcal{M}_{\rm eff} =\nonumber \\
&&\hspace*{-0.5cm}{\small - \left( \begin{array}{cc}
    M_D M_R^{-1} M^T_D + M_D M^{-1} M^T_L +M_L M^{-1} M_D^T &  M_D M_R^{-1} M^T    \\
    M M_R^{-1} M_D^T       & M M_R^{-1} M^T  
        \end{array} \right)
        }\nonumber
\label{eqn:eff_numatrix}       
\end{eqnarray}
An unique feature of extended linear seesaw mechanism, with desired mass hierarchy 
$M_R > M, M_D \gg M_L,\mu$ is that the type-I seesaw contribution to light neutrino masses is 
exactly canceled out, and the mass formula for light neutrinos (with $\mu=0$) becomes
\begin{eqnarray}
m_\nu=-M_D M^{-1} M^T_L - M_L M^{-1} M_D^T\, .
\end{eqnarray}
The mass formula for heavy RH neutrinos and heavy sterile neutrinos are
\bea
M_{S}& \sim & - M M^{-1}_R M^T   \\ 
M_{N} & \sim &  M_R +\cdots \, ,
\label{eq:mass}
\eea
For a case study, we have considered the model parameters as presented in Table.\ref{tab:constraint}.
\begin{table}[h!]
\centering
\begin{tabular}{c|c||c|c||c}
\hline 
$M_D$     & $M$          & $\theta_{S\nu} \simeq M_D/M$  & $M_R$         & $|M_S| \simeq M^2/M_R$ \\ 
\hline \hline
$1$ GeV      & $10^{5}$ GeV & $10^{-5}$  GeV      & $10^{15}$ GeV &  $7$ keV  \\ 
$0.1$ GeV    & $10^{4}$ GeV & $10^{-5}$  GeV      & $10^{13}$ GeV &  $7$ keV  \\
$0.01$ GeV   & $10^{3}$ GeV & $10^{-5}$  GeV      & $10^{11}$ GeV &  $7$ keV  \\
$0.0001$ GeV & $10^{2}$ GeV & $10^{-6}$  GeV      & $10^{9}$ GeV  &  $7$ keV  \\
\hline  
\end{tabular}
\caption{Input model parameters within extended seesaw mechanism examined 
         for one generation case in order to yield $7$ keV sterile neutrino 
         dark matter mass and sterile-active neutrino mixing of the order of 
         $10^{-5}$.}
\label{tab:constraint}
\end{table}
Using these model parameters one can find the light neutrino mass through extended linear seesaw contribution 
to be $m_\nu=(M_D/M) M_L=0.1$ eV for $10$ keV mass for $M_L$.

As seen from the Table.\ref{tab:constraint}, the extended linear seesaw mechanism 
consistently results $7$ keV lightest sterile neutrino dark matter for all choices 
of the model parameters and the active-sterile neutrino mixing at order 
of $10^{-5}$. Thus, the framework not only holds an answer for the $3.5$ keV X-ray 
line as observed by XMM-Newton X-ray Space observatory but also agrees  
with the oscillation data.

\section{Numerical Results:}
We present here a numerical study of masses and mixing within extended linear seesaw framework 
with $\nu_i, N_i, S_i\,$ i=1,2,3 where the model parameters may have Left-Right symmetric origin 
or $SO(10)$ origin. In such extended seesaw scheme 
both the sterile neutrinos have Majorana mass (where $S$ has mass of the order $M^2/M_R$ 
and $N$ has mass of the order $M_R$) and their admixture with the ordinary neutrinos is 
suppressed by the factors, $M_D/M$ and $M_D/M_R$ for $S-\nu$ and $N-\nu$ cases, respectively.  
respectively. Contrary to the case of canonical seesaw where the GUT scale right-handed 
neutrino mass (equivalent to seesaw scale) makes the model inaccessible to 
current accelerator experiments, here the mass of $M$ can be much smaller since the light neutrino 
mass formula involves $(M_D/M)$, $M_L$ and the sub-eV scale of light neutrinos can 
be consistently addressed by suitably adjusting the small lepton number violating term $M_L$ 
even with EW scale of $M$.

The Dirac-neutrino mass matrix is an arbitrary complex matrix in the present case, but there is 
a possibility that $M_D$ can be similar to the charged lepton mass matrix $M_\ell$ ($M_D$ is similar 
to the up-type quark mass matrix $M_u$) originating from left-right model (high scale Pati-Salam 
symmetry or $SO(10)$ GUT) \cite{left-right-group}. The other ${\small N-S}$ mixing mass matrix $M$, in principle, can take 
any form, but we have considered it to be diagonal for the sake of simplicity. However, the 
diagonal elements can be constrained from the existing experimental bound on unitarity violation 
in the lepton sector. Similarly, we consider here the diagonal structure for $M_R$ though both 
diagonal and general form of $M_R$ is possible when this extended seesaw framework will be embedded 
in a $SO(10)$ GUT model. 

\noindent
{\bf \, A.\,\,$M_D \simeq M_{u}$ signifying its origin of high scale Pati-Salam symmetry or $SO(10)$ GUT:~}

Within high scale Pati-Salam symmetry relating quarks and leptons with each other or in $SO(10)$ GUT models, 
it is found that the Dirac neutrino mass matrix is similar to the up-type quark mass matrix. Using the running 
masses $(m_u, m_c, m_t)=(0.00233, 1.275, 160)$~GeV and Cabbibo-Kobayashi-Maskawa mixing matrix, $V_{CKM}$ \cite{PDG}, 
the Dirac neutrino mass matrix is structured to be,
\bea 
&&M_D(\rm GeV)\simeq M_u=V_{CKM}\hat{M}_uV^T_{CKM} \nonumber\\
&&\hspace*{-0.6cm}=
\begin{pmatrix}
0.06         &  0.3-0.02i  &  0.55-0.53i\\
0.30-0.02i  &  1.48-0.0i     &  6.53-0.001i\\
0.55-0.5i    &  6.534-0.0009i &  159.7
\end{pmatrix}.
\eea

Using this Dirac neutrino mass matrix, diagonal structure of $N-S$ mixing matrix 
$M=\mbox{diag}\left(8.4 \times 10^{3}, 10^{5}, 10^{6} \right)\,$ GeV, and heaviest 
RH neutrino mass matrix $M_R=\mbox{diag}\left( 10^{13}, 2 \times 10^{13}, 5 \times 
10^{13} \right)\,$ GeV, we can derive the sterile-active neutrino mass matrix as 
\begin{eqnarray}
&&\widehat{M}_S=\left(7\, \mbox{keV}, 0.5\,\mbox{MeV}, 20\,\mbox{MeV} \right)\,  \nonumber \\
&& \hspace*{-0.4cm}
\theta_{\nu\,s}=10^{-6} \times
\begin{pmatrix}
7.94                 & 0.302-0.22 i  & 0.55-0.53 i  \\
35.3-2.62 i          & 14.8          & 6.53-0.001 i  \\
65.4-63.1 i          & 65.3-0.009    & 159
\end{pmatrix}\nonumber \\
\end{eqnarray}

Notably, the presence of a large number of intermediate symmetry breaking steps from $SO(10)$ to SM 
affects the value of $M_D\simeq M_u$ \cite{inverse, patra-jhep} .
As discussed in Ref. \cite{patra-jhep},
a particular form of $M_D$ including RG Corrections is 
\begin{eqnarray}
\label{eq:md_with_rge}
&&M_D = \nonumber \\
&&\hspace*{-0.5cm}
\begin{pmatrix}
 0.022          & 0.098-0.016 i    & 0.146-0.385 i\\
 0.098+0.016 i  & 0.6319           & 4.88+0.0003 i\\
 0.146+0.385 i  & 4.884-0.0003 i   & 117.8
\end{pmatrix}\text{GeV}.\nonumber \\
\end{eqnarray}
Assuming $M$ to be diagonal for the sake of simplicity, $M\equiv{\rm diag}(M_1, M_2, M_3) 
\simeq \mbox{diag}(3\times 10^{3}, 10^{5}, 5 \times 10^{5})$ GeV and 
heaviest RH neutrino mass matrix $M_R \equiv{\rm diag}(M_{R_1}, M_{R_2}, M_{R_3}) 
\simeq \mbox{diag}(1.25\times 10^{12}, 2\times 10^{13}, 5 \times 10^{13})$ GeV, the sterile 
neutrino mass eigenvalues for $S$ is found to be 
$\widehat{M}_S=\left(7\, \mbox{keV}, 0.5\,\mbox{MeV}, 12.5\,\mbox{MeV} \right)\,$ 
and corresponding sterile-active neutrino mixing is of the order $10^{-6}$.

\noindent \\
{\bf \, B.\,\,Dirac neutrino mass matrix is similar to charged lepton mass matrix ($M_D \simeq M_\ell$):\,} 

The Dirac neutrino mass matrix can be similar to charged lepton mass matrix 
as expected from Left-Right symmetric theory \cite{left-right-group} and when we are working in a basis where 
charged lepton mass matrix is already diagonal it becomes, 
$M_D\simeq M_\ell = \mbox{diag}(m_e, m_\mu, m_\tau)$. Assuming diagonal structure of 
$M=\mbox{diag}(50, 10000,100000)$ GeV and $|M_R| \geq 3.35 \times 10^{8}$ 
GeV (inverted pattern of heavy neutrinos mass matrix $M_R$ of any general form) 
along with $M_D\simeq M_\ell$ 
the estimated value of the sterile neutrino mass eigenvalues becomes
$\widehat{M}_S=\left(7\, \mbox{keV}, 0.5\,\mbox{MeV}, \mbox{2\,GeV},\, \mbox{50\, GeV} \right)\,$ 
and corresponding sterile-active neutrino mixing of the order $10^{-6}$. 
\section{3.51\, keV X-ray line signal}
\begin{figure}[h!]
\centering
\includegraphics[scale=0.6,angle=0]{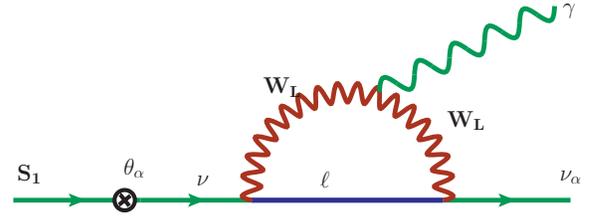}
\caption{Feynman diagram for two body radiative decay of sterile neutrino dark matter 
         into a ordinary neutrino plus a monochromatic photon line with energy 
         $3.51$ keV.}
\label{feyn:nugamma}
\end{figure}

Sterile neutrinos should mix slightly with active neutrinos in order to rationalize their origin in the 
early universe. Possibly, this mixing causes the active neutrinos to oscillate to sterile ones which further 
increases the flock of sterile neutrinos and backs the relic abundance of DM. In the nick of time, the same mixing 
causes a sterile neutrino ($S_1$) to decay into a light active neutrino plus a monochromatic photon line 
of energy $E_\gamma=m_S/2$. The analytic formula for the decay width of this process as presented 
in Fig.\ref{feyn:nugamma} is,
\begin{eqnarray}
&&\Gamma_{S_1 \to \nu \gamma} = \frac{9 \alpha G^2_F}{1024 \pi^2} \sin^22\theta\, m^5_{S}\, ,
\end{eqnarray}
where $\alpha$ is the fine structure constant for electromagnetic interaction, $G_F$ is the universal Fermi 
coupling constant, $m_{S}\simeq M_{DM}$ is the mass of the sterile neutrino dark matter. 

\begin{figure}[h!]
\centering
\includegraphics[scale=0.6,angle=0]{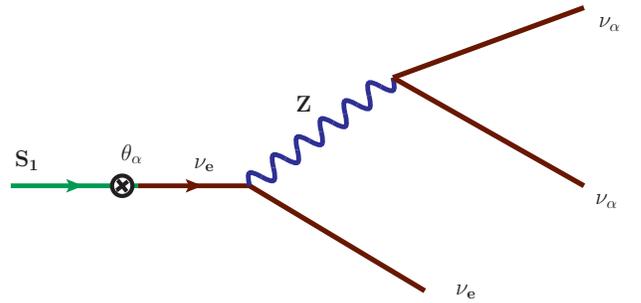}
\caption{Feynman diagram for sterile neutrino decay $S \to \nu_e \nu_\alpha \overline{\nu_\alpha}$ 
         via neutral current interactions. }
\label{feyn:3nu}
\end{figure}

Elseways, the sterile neutrino decays into three ordinary neutrinos as shown in 
Fig.\ref{feyn:3nu} with the decay rate,
\begin{eqnarray}
&&\Gamma_{S_1 \to \nu \nu \nu} = \frac{4 G^2_F}{384 \pi^2} \sin^2(2\theta) m^5_{S}\, ,
\end{eqnarray}

For observed $3.5$ keV X-ray line signal, it is calculated that the decay rate for sterile 
neutrino dark matter to $\nu$ and a photon is $\sim 3.3\times 10^{-52}$ GeV. Using the PDG 
values $\alpha_{\rm em}=1/127.94$, $G_F=1.1662 \times 10^{19}$ GeV and sterile-active neutrino 
mixing of the order of $10^{-10}$, the dominant decay mode $S \to 3 \nu$ with a keV mass 
sterile neutrino dark matter results,
\begin{eqnarray}
\Gamma_{S_1\to 3\nu} \simeq 1.43 \times 10^{-53}\, \left(\frac{\sin^2 2\theta}{10^{-10}} \right) 
                     \left(\frac{m_{S_1}}{\mbox{keV}} \right)^{5} \mbox{GeV}\, ,
\end{eqnarray}
while for radiative two body decay yields 
\begin{eqnarray}
\Gamma_{S_1\to \nu + \gamma} \simeq 9.47 \times 10^{-56}\, \left(\frac{\sin^2 2\theta}{10^{-10}} \right) 
                     \left(\frac{m_{S_1}}{\mbox{keV}} \right)^{5} \mbox{GeV}\, .
\end{eqnarray}

\begin{figure}[h!]
\centering
\includegraphics[scale=0.9,angle=0]{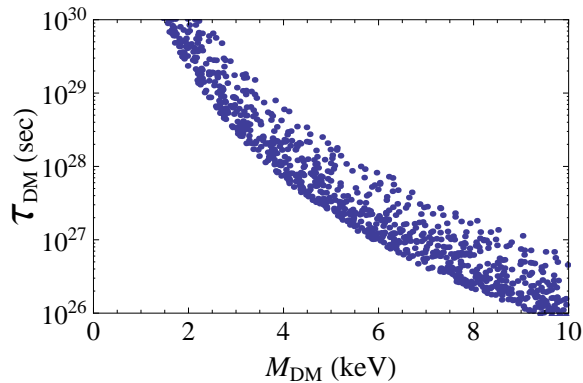}
\caption{Estimation of life-time of decaying dark matter ${\bf \tau}_{DM}$ wrt dark matter mass $M_{DM}$}
\label{scatter-plot}
\end{figure}
 
The mass and mixing of a sterile neutrino should be indeed tiny to make itself a viable DM candidate which survives 
much longer than the universe. We demonstrate here how the stability of the sterile neutrino dark matter 
of $7$ keV is fulfilled with a life-time $\tau_{DM}\simeq 6.2\times 10^{27}$ secs as shown in 
Fig.\,\ref{scatter-plot}. With $\sin^2 2\theta \simeq 
10^{-10}$ and $7$ keV sterile neutrino mass, the life-time for 3-body three level decay can be written as
\begin{eqnarray}
\tau_{S_1\to 3\nu} \simeq 4.5 \times 10^{28}\, \left(\frac{10^{-10}}{\sin^2 2\theta} \right) 
                     \left(\frac{\mbox{keV}}{m_{S_1}} \right)^{5} \mbox{sec}\, ,
\end{eqnarray}
and that for a radiative 2-body decay as
\begin{eqnarray}
\tau_{S_1\to 3\nu} \simeq 6.9 \times 10^{30}\, \left(\frac{10^{-10}}{\sin^2 2\theta} \right) 
                     \left(\frac{\mbox{keV}}{m_{S_1}} \right)^{5} \mbox{sec}\, ,
\end{eqnarray}
which assures us that a sterile neutrino dark matter with mass $m_S=2 E_\gamma= 7 $ keV 
can easily satisfy the stability criteria. Some other experiments on X-ray signal with different energy (around keV)  
put a constraint on sterile-active neutrino mixing as \cite{xray-constraints},
\begin{eqnarray}
\sin^2 2 \theta \leq 1.2 \times 10^{-5} \left( \frac{m_{S_1}}{\rm keV}\right)^{-5}
\end{eqnarray}

At present, predicting the exact relic density of dark matter is another necessity of a DM model. 
In case of cold dark matter, it is assumed that its relic abundance increases by thermal freezing-out mechanism 
based on the thermally averaged effective DM annihilation cross section at the time of freeze-out. 
However, relic abundance of keV mass dark matter is supported by specific production mechanisms 
 which involve more engaging calculations. One such mechanism is Dodelson-Widrow mechanism \cite{dodelson} which 
provides the formula for relic density of DM as,
\begin{eqnarray}
\Omega_S h^2 \simeq 0.3 \left(\frac{\sin^2 2\theta}{10^{-10}} \right) 
                     \left(\frac{m_{S_1}}{\mbox{100\,keV}} \right)^{2},
\end{eqnarray}
Few more works have also determined the relic abundance of a 10 keV sterile neutrino 
DM to be,
\begin{eqnarray}
\Omega_S h^2 \simeq 10^{-7} d_\alpha \left(\frac{\sin^2 2\theta}{10^{-10}} \right) 
                     \left(\frac{m_{S_1}}{\mbox{10\,keV}} \right) 
                     \left(\frac{T_R}{\mbox{5\,MeV}} \right)^3\, ,
\end{eqnarray}
with a presupposition that the universe after being inflated never attained a temperature above few MeV 
\cite{DM-relic-a,DM-relic-b,DM-relic-c}. 

But with the above relations it seems clear that for both the cases the allowed range of $\sin^2 2\theta$ and 
$m_S$ (from the recent X-ray observation)  hardly validates the relic abundance of sterile neutrinos. 
Whereas, in order to justify the fact that dark matter (DM) contributes $25\%$ to the total 
energy density of the present universe, its relic abundance should be $\Omega_{DM} h^2 = 0.119$. 
In this regard, the Shi-Fuller mechanism \cite{shi-fuller} predicts an acceptable value with a 
$7$ keV sterile neutrino dark matter. In addition to this, one may also go through refs.\cite{relic,relic-correct} 
for the exact calculations relating to relic abundance. 



\section{Conclusion}
We have shown how a $7$ keV sterile neutrino dark matter and the corresponding sterile-active mixing angle 
$\simeq 10^{-5}$ can be accommodated successfully in an extended seesaw framework with simple 
addition of two types of sterile fermions to the minimal particle content of Standard Model. 
The radiative decay of the $7$ keV sterile neutrino dark matter into the SM light neutrino plus 
a monochromatic photon can easily explain the recent unidentified $3.5$ keV X-ray line signal 
as observed by XMM-Newton Telescope. The numerical results of the framed model stands parallel 
to the oscillation data explaining the sub-eV scale of light neutrino masses. 
More significant a point; this framework having high scale heavy neutrinos with masses 
$> 10^{9}$ GeV can ably answer another vital issue of particle physics called 
matter-antimatter asymmetry of our present Universe.

\section{Acknowledgement}
Prativa Pritimita is grateful to the Department of Science and Technology, Govt. of India for INSPIRE 
Fellowship (IF140299). The work of Sudhanwa Patra is partly supported by the Department of Science and Technology, 
Govt. of India under the financial grant SERB/F/482/2014-15. We are sincerely thankful to Sourov Roy for going through 
the manuscript and giving his valuable remarks.

\end{document}